# JCGM 101-COMPLIANT UNCERTAINTY EVALUATION USING VIRTUAL EXPERIMENTS

*Finn Hughes* [a, *], *Manuel Marschall* [a], *Gerd Wübbeler* [a], *Gertjan Kok* [b], *Marcel van Dijk* [b], *Clemens Elster* [a]

[a] Physikalisch-Technische Bundesanstalt (PTB), Abbestraße 2-12, 10587 Berlin, Germany, finn.hughes@ptb.de
[b] VSL Dutch Metrology Institute, 2629 JA Delft, The Netherlands
* Corresponding author

*Abstract* − Virtual experiments (VEs), a modern tool in metrology, can be used to help perform an uncertainty evaluation for the measurand. Current guidelines in metrology do not cover the many possibilities to incorporate VEs into an uncertainty evaluation, and it is often difficult to assess if the intended use of a VE complies with said guidelines. In recent work, it was shown that a VE can be used in conjunction with real measurement data and a Monte Carlo procedure to produce equal results to a supplement of the *Guide to the Expression of Uncertainty in Measurement*. However, this was shown only for linear measurement models. In this work, we extend this Monte Carlo approach to a common class of non-linear measurement models and more complex VEs, providing a reference approach for suitable uncertainty evaluations involving VEs. Numerical examples are given to show that the theoretical derivations hold in a practical scenario.

## 1. INTRODUCTION

The typical application of virtual experiments (VEs) in a metrological context is for simulating data for a sensitivity analysis regarding the uncertainty sources involved in the VE [1], or to perform an uncertainty evaluation for the measurand when the collection process for real data is complex or costly [2]. With an uncertainty evaluation, to ensure that the calculated uncertainties are reliable, it is important that an appropriate procedure is followed. As the prominence, knowledge, and importance of VEs in research rapidly increases, the need to sufficiently regulate their usage in metrological applications emerges in parallel.

Much literature is available on how to conduct an uncertainty evaluation. The bedrock for uncertainty evaluation protocol in metrology is provided in the *Guide to the Expression of Uncertainty in Measurement* (JCGM 100) [3] and its supplements, in particular JCGM 101 [4]. In more recent times, extensions to uncertainty evaluations involving VEs have been discussed in the context of metrological applications. For example, [5] and [6] provide examples where VEs have been used in an uncertainty evaluation involving a coordinate measuring machine (CMM), while [7] presents how VEs can be adequately used in scatterometry. Moreover, [5] provides a Monte Carlo sampling procedure using a VE that equals the results of the JCGM 101 approach for linear measurement models and hence linear VEs.

There are however issues with JCGM 101 compliancy in uncertainty evaluations using VEs. For instance, [6] highlights some of the issues arising when using a VE, e.g., without the true measurand or consideration on the sensitivity to input values. In [5], the distinction between a measurement model for the measurand as required by JCGM 101 and the model underlying a VE is illustrated. Guidance on the appropriate implementation of a VE in an uncertainty evaluation is therefore paramount to scientific development.

This paper will demonstrate how VEs can be used in combination with Monte Carlo sampling to conduct an uncertainty evaluation for a wide class of non-linear measurement models, which yields equivalent outcomes to those obtained through the JCGM 101 framework.

After specifying what constitutes a VE in section 2 and how it relates to a statistical model, the approach for incorporating a VE in a JCGM 101-compliant uncertainty evaluation is proposed in section 3. Section 4 then illustrates practical examples, both generic and specific to a metrological application, to demonstrate the JCGM 101 compliance of the developed procedure. A conclusion can be found in section 5, along with suggestions on potential future directions of research.

## 2. VIRTUAL EXPERIMENTS

A 'measurement model', the starting point of a JCGM 101-compliant uncertainty evaluation in metrology, consists of a function relating the input quantity representing the information contained in the measurement data, along with other input quantities which should cover all significant uncertainty sources, to the actual measurand or quantity of interest. On the other hand, in a VE, virtual measurement data can be simulated using a forward model involving a pre-determined value for the measurand, as well as some additional probabilistic elements. VEs are of use due to their ability to efficiently model expensive experiments and generate much more data than could be obtained from real measurements. This also makes VEs generally cost-effective and efficient. Current literature (e.g. [5], [6], [7]) presents (1) as an example of a VE, with function $g$ representing the forward model and the additional random fluctuations modelled by some simple distribution such as $\varepsilon \sim N(0, \sigma^2)$ (that is, $\varepsilon$ follows a standard Gaussian distribution with mean 0 and variance $\sigma^2$):

$$x^{ve} = g(Y, Z) + \varepsilon. \qquad (1)$$

Simulations following (1) will produce virtual data $x^{ve}$ through realisations of $\varepsilon$ and specification of fixed values for the measurand $Y$ and the additional input variables $Z$. Let $X$ denote the quantity associated with the data. In the following we will call the assumed scalar $X$ a quantity with Type A information, and the (possibly multivariate) $Z$ a quantity with Type B information. The random fluctuations shall mimic the variability observed in repeated real measurement results.

Model (1) requires somewhat sharp knowledge about the distributional form of the random noise and would hence be less appropriate in the circumstance where the random fluctuations produced by a comprehensive VE are no longer separable from the forward model. Although it is often assumed that the Gaussian noise covers the statistical deviations of repeated measurements, in practice the behaviour is often not so simple. Indeed, when simulating data with the same input values in a VE, one would still expect the outputs to differ in general as a result of the additional probabilistic elements. An adapted model should thus attempt to model the fluctuations of repeated simulations of the VE in a more realistic way, e.g. as random fluctuations within the forward model, resulting in non-Gaussian distributions of the virtual data. Mathematically, this can be written as a statistical model of the form:

$$x^{ve} = G(Y, Z, w), w \sim F_v(W). \quad (2)$$

Within function $G$ in (2), one considers $Y$ and $Z$ (defined as before) as fixed parameters since they are also assigned to fixed quantities in repeated real experiments. The random fluctuations in the outcomes are accounted for by the distribution of $W$, denoted $F_v(W)$, where $v$ represents the additional parameters of said distribution. The forward model from (1) can still be deduced by simply equating $W$ to 0.

The important distinction here is that the VE in general should not be equated to only the forward model – a point which is arguably not proclaimed distinctly enough in literature. The impact of a more realistic virtual model of random fluctuations observed in repeated real data will be considered elsewhere. In the following we will develop a JCGM 101-compliant procedure for a set of observed data using the output of a VE under the simplified assumption of Gaussian fluctuations in real data, and we will assume knowledge about the variance $\sigma^2$ of said fluctuations.

## 3. METHODS

The task addressed in this section is when given a VE in the form of (1), a distribution of the Type B variable $Z$, and a set of real observations $x_1, \ldots, x_m$ for the Type A variable $X$, design a Monte Carlo procedure using repeated applications of the VE which produces samples from a distribution obtainable if an equivalent measurement model was available and a JCGM 101 approach was applied instead. Due to the assumption of a Gaussian distribution with known variance for the observed data, the data can be summarised in the empirical mean $\bar{x} = \frac{1}{m}\sum_{i=1}^{m} x_i$, which will be taken as an input for the considered uncertainty evaluations.

After providing an outline of uncertainty evaluation procedures from the current literature, such as the JCGM 101 approach and the VE and Monte Carlo combination for linear models from [5] (referred to in this paper as the MC-VE approach), an extension to [5] is proposed, such that the procedure is also applicable to a wide class of non-linear models.

### 3.1. JCGM 101 approach

At first, it is important to outline some of the established requirements for a JCGM 101 compliant uncertainty evaluation. To comply with JCGM 101, an uncertainty evaluation must follow the step procedure outlined below:

1. Define the measurement model: $Y = f(X, Z)$.
2. Assign a PDF to the input quantities $X$ and $Z$.
3. Apply the Monte Carlo propagation method.
4. Summarise the results.

Additional alterations, such as adding contributions for systematic effects, following the completion of step 4 would constitute a failure of JCGM 101 compliancy, and hence deem it an invalid uncertainty evaluation under the regulations defined in JCGM 101. Deviating from this step procedure would also generally render an uncertainty evaluation non-compliant. However, a valid uncertainty evaluation could still be reached provided the method is found to be equivalent to a JCGM 101-compliant uncertainty evaluation.

The original JCGM 101 document did not consider, or at least address specifically, how VEs could feature within an uncertainty evaluation. Logically, if the VE is used within the step procedure, there should be no issue with regards to complying with JCGM 101. Our approach developed later in this section uses the VE in steps 2 or 3, so would hence pass this requirement.

To follow the JCGM 101 uncertainty evaluation approach, the required knowledge must include a known distribution for the Type B variables $z \sim \pi(Z)$ and a measurement model of the form $Y = f(X, Z)$. The application of the JCGM 101 approach then proceeds as outlined in Algorithm 1.

Algorithm 1
Given the mean $\bar{x}$ of $m$ observations from a Gaussian distribution with known variance $\sigma^2$, choose $n$ (e.g. $n = 10^6$).
For every $i = 1, \ldots, n$:
  1) Draw samples $x'_i \sim N\left(\bar{x}, \frac{\sigma^2}{m}\right)$, $z_i \sim \pi(Z)$.
  2) Calculate $y'_i = f(x'_i, z_i)$.
Use $y'_1, \ldots, y'_n$ to approximate $\pi(Y)$ and obtain an estimate of $Y$ and an associated uncertainty e.g. the mean and standard deviation.

### 3.2. MC-VE approach: linear models

When considering a linear VE, there has already been a successful procedure shown by [5] for replicating the results of the JCGM 101 uncertainty evaluation approach through repeatedly running the VE with an arbitrarily chosen value for the measurand and varying values for the Type B variable $Z$. A simple correction of the results, utilising the observed data $\bar{x}$ and the arbitrarily chosen value for the measurand, then yields a sample from a distribution which equals the distribution for the measurand obtained by the application of the JCGM 101 approach. For further details we refer the reader to [5].

This procedure is useful in that the sampled values do not depend on the hypothetical 'chosen' measurand value, which would negate issues with JCGM 101 compliancy surrounding the (general) inability to use the true measurand when generating simulations of the virtual data.

It is of course necessary to assume that the variance of the data distribution, $\sigma^2$, is known. For unknown variance, JCGM 101 compliancy would not be feasible as the JCGM 101 Monte Carlo approach will sample from a t-distribution instead, contrasting to the normally distributed random errors produced by the VE, and hence different distributions for the measurand would emerge [5].

### 3.3. MC-VE approach: non-linear models

While the linear MC-VE approach has already been established in previous literature [5], the procedure cannot be applied in the case where a non-linear VE is present instead. For this, we propose a class of VEs of the form (6), which can be used to generate the virtual data, assuming $\Delta_1(z) \neq 0$:

$$x_i^{ve} = \Delta_1(z_i) \cdot y + \Delta_2(z_i) + \varepsilon_i. \tag{3}$$

Here, $y$ is a value for the measurand, with $x_i^{ve}$ representing virtual realisations of $X$ and $z_i \sim \pi(Z)$ representing realisations of the state of knowledge distribution about the value of the Type B variable. $\Delta_1(z_i)$ and $\Delta_2(z_i)$ are arbitrary functions. Provided one has a set of $m$ real observations with mean $\bar{x}$ from a Gaussian distribution with known variance $\sigma^2$, along with the distribution $\pi(Z)$, the Monte Carlo method can be used to obtain samples for the measurand through the VE in a JCGM 101-compliant way.

Following the standard JCGM 101 approach, one would collect $n$ samples each from $x' \sim N(\bar{x}, \sigma^2/m)$ and $z \sim \pi(Z)$ to obtain samples $y_i'$ of the measurand using the following formula:

$$y_i' = \frac{x_i' - \Delta_2(z_i)}{\Delta_1(z_i)}, \tag{4}$$

which corresponds to the formula $y_i' = f(x_i', z_i)$ in Algorithm 1. It also follows that one can sample values $y_i'$ for the measurand conditional on the values of $z_i$ from the following Gaussian distribution:

$$y_i'|z_i \sim N\left(\frac{\bar{x} - \Delta_2(z_i)}{\Delta_1(z_i)}, \frac{\sigma^2}{m(\Delta_1(z_i))^2}\right). \tag{5}$$

This, however, requires explicit knowledge about all elements of the measurement model (7).

If one were to instead follow the approach involving the VE, one could sample the measurand through a Monte Carlo approach by first using an arbitrarily chosen value $y_0$ for the measurand to generate simulated values for each $z_i$:

$$\bar{x}_i^{ve} = \frac{1}{m}\sum_{j=1}^{m} x_{i,j}^{ve} = \frac{1}{m}\sum_{j=1}^{m}\left(\Delta_1(z_i) \cdot y_0 + \Delta_2(z_i) + \varepsilon_j\right). \tag{6}$$

These samples of the empirical mean of the virtual data for the Type A variable given the realisation of the Type B quantity $z_i$ follow a Gaussian distribution, namely $\bar{x}_i^{ve}|z_i \sim N(\Delta_1(z_i) \cdot y_0 + \Delta_2(z_i), \sigma^2/m)$. Furthermore, one can obtain samples of the measurand using the following formula:

$$y_i^{ve} = \frac{\bar{x} - \bar{x}_i^{ve}}{\Delta_1(z_i)} + y_0, \tag{7}$$

and it can be shown through the expansion of $\bar{x}_i^{ve}$ in (10), using the formula presented in (9), that the values $y_i^{ve}$ do not depend on the hypothetical value $y_0$ required to run the VE, while the conditional distribution given $z_i$ equals that obtained in (8) for the JCGM 101 approach. This feature mimics the approach developed in [5], but now it is applicable also for nonlinear models of the form (6).

Since the same distribution $\pi(Z)$ is used as for the JCGM 101 approach, and because the respective conditional distributions are the same for the MC-VE approach and the JCGM 101 approach, the samples produced by (10) are samples from the distribution for the measurand obtained by the JCGM 101 Monte Carlo procedure.

Algorithm 2 summarises this JCGM 101-compliant MC-VE procedure for a specific class of non-linear models.

Algorithm 2
Given the mean $\bar{x}$ of $m$ observations from a Gaussian distribution with known variance $\sigma^2$, choose $n$ (e.g. $n = 10^6$) and some value $y_0$ for the virtual measurand.
For every $i = 1, \ldots, n$:
   1) Draw samples $z_i \sim \pi(Z)$ and $\varepsilon_{i,j} \sim N(0, \sigma^2)$ for $j = 1, \ldots, m$.
   2) Calculate simulated observations $x_{i,1}^{ve}, \ldots, x_{i,m}^{ve}$ through $m$ runs of the VE in model (6), while using the chosen virtual measurand $y_0$ and $z_i$ and $\varepsilon_{i,j}$ generated in step 1).
   3) Calculate the mean of the simulated observations $\bar{x}_i^{ve} = \frac{1}{m}\sum_{j=1}^{m} x_{i,j}^{ve}$.
   4) Calculate $y_i^{ve} = \frac{\bar{x} - \bar{x}_i^{ve}}{\Delta_1(z_i)} + y_0$.
Use $y_1^{ve}, \ldots, y_n^{ve}$ to approximate $\pi(Y)$ and obtain an estimate of $Y$ and an associated uncertainty.

An advantage of using Algorithm 2, compared to the classical JCGM 101 Algorithm 1, is that the VE can be used as a black-box model, meaning one does not need to identify the additive, possibly non-linear effect $\Delta_2(z_i)$ and may simply evaluate the VE and its derivative with respect to $Y$ repeatedly. Note that, when considering non-linear VE models, sampling estimates of the measurand requires knowledge of the partial derivative of (6) with respect to the measurand i.e. knowledge of $\Delta_1(z_i)$ - an additional requirement compared to the MC-VE approach for linear models. This derivative should generally be obtainable given that there should be sufficient knowledge of the model parameters available, for instance using tools from automatic differentiation, backpropagation, or numerical difference quotients. When the gradient is calculated numerically, possibly utilising a complicated numerical VE, one would need to check that the VE actually falls into the class of non-linear models considered in (6), otherwise Algorithm 2 would yield only an approximation.

The extension to the procedure in [5] allows models of a wide class of non-linearity to have an uncertainty evaluation conducted through the Monte Carlo sampling approach involving the VE. The use here is that we now have provided a reference procedure in line with JCGM 101 compliancy, which should ensure valid uncertainty evaluations are always feasible for a valid VE model. Similarly to the linear case, knowledge about the variance of the data distribution in the corresponding real experiment is, however, required.

## 4. PRACTICAL APPLICATIONS

We conducted a series of applications, namely a generic numerical example, and a real-world example based on mass calibration. Both of our applications make use of the class of non-linear models considered in section 3.3. The examples illustrate the possibility to utilise a VE for a JCGM 101-compliant uncertainty evaluation.

### 4.1. Generic example

In this section, we adopt an example featured in [5]. Here, using the form presented in (6), we set $\Delta_1(z) = 1 + z$ and $\Delta_2(z) = 0$:

$$x = (1 + z) \cdot y + \varepsilon. \qquad (8)$$

$Z$ follows a uniform distribution with lower bound 5 and upper bound 10. It is assumed that the additional fluctuations $\varepsilon$ follow a standard zero-mean Gaussian distribution with known variance $\sigma^2 = 1$. A single measurement $\bar{x} = 50$ was taken. The VE in this case simply involves evaluating model (6) with these given specifications and an arbitrary choice for the value of the measurand $y_0$.

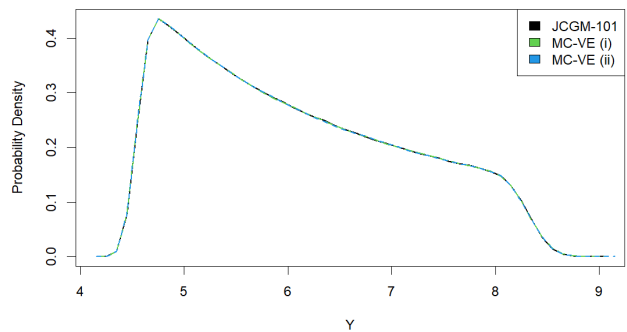

Figure 1. A density plot for the samples obtained through each of the respective procedures for the non-linear model $x = y(z + 1) + \varepsilon$. MC-VE (i) denotes the MC-VE approach with the chosen measurand value $y_0 = 50/8.5$, while MC-VE (ii) uses the value $y_0 = -100$.

Figure 1 displays the distribution when applying JCGM 101 using $n = 10^7$ Monte Carlo runs. Also shown is the distribution obtained by the proposed MC-VE approach, cf. section 3.3. In MC-VE (i), the hypothetical value $y_0 = 50/8.5$ for the measurand has been taken when carrying out the VE. To demonstrate the lack of dependence on the chosen value for the measurand in the VE, a third distribution (labelled MC-VE (ii)) is produced by again applying the MC-VE approach but now using $y_0 = -100$ as a hypothetical value for the measurand. All three results are found to be equivalent in Figure 1 and the summary statistics listed in Table 1 – remaining discrepancies are due to the random fluctuations inherent in the calculated distributions.

Table 1. Summary statistics for the samples obtained through each of the respective procedures for the non-linear model $x = y(z + 1) + \varepsilon$. MC-VE (i) denotes the MC-VE approach with the chosen measurand value $y_0 = 50/8.5$, while MC-VE (ii) uses the value $y_0 = -100$. Each statistic is given to 3 significant figures.

| Statistic | JCGM 101 | MC-VE (i) | MC-VE (ii) |
|---|---|---|---|
| Mean | 6.06 | 6.06 | 6.06 |
| Standard Deviation | 1.07 | 1.07 | 1.07 |
| Shortest 95% CI | [4.50,8.06] | [4.50,8.06] | [4.50,8.06] |

### 4.2. Real world example

To demonstrate the equivalence holds beyond a generic example, we can use the respective approaches in a 'real world' case. For instance, we have considered the mass calibration example from JCGM 101 and [8], which assigns the measurand $Y$ to be the deviation of the conventional mass of a (hypothetical) weight from the nominal mass, $m_{nom}$. The VE in this example is given by:

$$X = \frac{Y - m_{nom}}{\left(1 + (\rho_a - 1.2)\left(\frac{1}{\rho_W} - \frac{1}{\rho_R}\right)\right)} - m_{R,c} + \varepsilon. \qquad (9)$$

Here, $m_{R,c}$ represents the conventional mass of reference weight $R$, $\rho_a$ represents the mass density of the air in which the experiment is operating in, and $\rho_W$ and $\rho_R$ are the mass densities of weight $W$ and reference weight $R$ respectively. $X$ represents the deviation of the conventional mass of $R$ from the nominal mass, which is a constant $m_{nom} = 10^5 \text{mg}$. Although $Z = (m_{R,c}, \rho_a, \rho_W, \rho_R)$ in this example has multiple parameters, the measurand $Y$ is still a univariate quantity. Hence, our proposed method is easily transferable to this example. While parameter $m_{R,c}$ follows a Gaussian distribution, $\rho_a$, $\rho_W$, and $\rho_R$ all follow independent uniform distributions. The exact definitions of the input quantities are given or inferred in [8] and copied into Table 2. We have assumed a known variance $\sigma^2 = 0.001 \text{mg}^2$ for the distribution of $X$ and that there were 5 observations with $\bar{x} = 1.2345$. For the MC-VE approach, the arbitrarily chosen measurand value was $y_0 = 1$.

Table 2. Input quantities used in the mass calibration example based on the example in [8], along with their units.

| Quantity | Unit | Distribution/Value |
|---|---|---|
| $y_0$ | mg | Constant (1) |
| $m_{R,c}$ | mg | $N(10^5, 0.0025)$ |
| $\rho_a$ | kg/m$^3$ | U(1.1,1.3) |
| $\rho_W$ | kg/m$^3$ | U(7000,9000) |
| $\rho_R$ | kg/m$^3$ | U(7950,8050) |

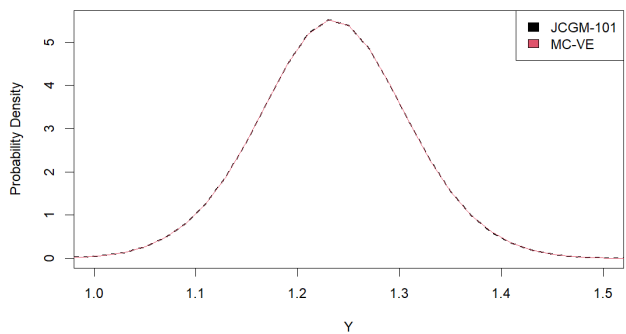

Figure 2. A density plot for the samples obtained through each of the respective procedures for the mass calibration example.

Again, after applying $10^7$ runs for both approaches, one can see the equivalence between the results in the density plot in Figure 2 and the statistics in Table 3 which hence reinforces that our proposed MC-VE approach for non-linear models of the form displayed in (6) complies with the JCGM 101 framework.

Table 3. Summary statistics for the samples obtained through each of the respective procedures for the mass calibration example from [8]. Each statistic is given to 3 significant figures in mg.

| Statistic | JCGM 101 | MC-VE |
|---|---|---|
| Mean | 1.23 | 1.23 |
| Standard Deviation | 0.0741 | 0.0741 |
| Shortest 95% CI | [1.09, 1.38] | [1.09, 1.38] |

## 5. CONCLUSION AND OUTLOOK

We have exhibited how a VE in combination with Monte Carlo sampling can be used in an uncertainty evaluation for non-linear models equivalently to a standard JCGM 101 approach. This is useful in that it ensures the existence of a reference procedure for how to conduct an appropriate uncertainty evaluation in the event that a VE should be used, regardless of the linearity of the VE model. The equivalence is valid for non-linear models falling into the considered, wide class of non-linear models established in (6). Cases where the developed procedure is particularly relevant are scenarios in which a VE is available but not the measurement model required for a direct application of JCGM 101.

Future research could focus on alternative choices of the distribution of the random noise. While in each reference procedure the added random fluctuations are assumed to follow a standard Gaussian distribution, it would be interesting to observe the implications of altering how such noise is modelled. This could potentially be through changing the distribution to another member of the location-scale family. Alternatively, while the current random errors are included as an additional term, under the defined VE function $G$ in section 2, there is the possibility of treating the noise as a product with the forward model, or as an additional function itself.

Future research could also include a statistical assessment for the compliance of the constructed methods to achieve explicit guidance in practical scenarios. That is, a hypothesis test or credible interval could perhaps be formulated to ensure that the virtual data do not differ significantly from their real, true counterparts.

## ACKNOWLEDGMENTS

This work was supported by the European Partnership on Metrology, co-financed from the European Union's Horizon Europe Research and Innovation Programme and by the Participating State [grant number 22DIT01 ViDiT].